\documentstyle[12pt]{article}
\begin{document}

\begin{center} {\large\bf Wigner's friend in context.}\\[5mm]

{R.S. Mackintosh}\\[5 mm]
{\it School of Physical Sciences, The Open University, Milton Keynes, MK7 6AA, UK}\\[5mm]
\end{center}

Email: raymond.mackintosh@open.ac.uk; Orcid: 0000-0002-2040-6193\\

{\bf Abstract:} {Wigner's friend is commonly invoked in discussion of quantum mechanics and its interpretation.. but who is the friend really? The publishing context of Wigner's article is not widely appreciated.  A recent paper by Ballentine is relevant.}

\begin{center}\today\end{center}
We are introduced to his friend by Wigner in his contribution to a book~\cite{wigner} which has the subtitle `An anthology of partly-baked ideas'. This anthology was collected and edited by the distinguished statistician (and computer pioneer) Irving J. Good.\footnote{Wikipedia gives the publication date as 1965 with publisher Capricorn Books. I purchased my copy in 1963 or 1964, and it was published in 1962.} The article, \emph{Remarks on the Mind-Body Question}, has been reprinted in Ref.~\cite{QTandM} and in Ref.~\cite{SandR} but neither of these mentions the subtitle of the original book. While one shouldn't judge an article just by the company it keeps, it is certainly interesting that the same collection  includes Harlow Shapley's `Concerning Life on Stellar Surfaces' and Marvin Minsky's `Winking at Computers.' Browsing the list of 123 contributions, it is obvious that many contributors took the challenge of producing something worthy of the book's subtitle very seriously. Maybe Wigner did too.

It is well known that, at one time, Wigner's resolution of the measurement problem of quantum mechanics involved human consciousness. It is also well known that in due course Wigner abandoned the idea that human consciousness could solve the measurement problem. Wigner's famous article, written in 1962, Ref.~\cite{WignerAJP} (also included in the article
collections, Refs.~\cite{QTandM,SandR}) does \emph{not} call upon human consciousness, but leaves the measurement problem as\ldots a problem.

Wigner\cite{wigner} quotes Heisenberg who, in an article of 1958 wrote: `The conception of objective reality evaporated \ldots into the mathematics which represents \ldots our knowledge of the behavior [of elementary particles]'. However,  on the other hand,  Heisenberg also wrote~\cite{heisenbergtoNB}: `The criticism of the Copenhagen interpretation rests \ldots on the anxiety that \ldots the concept of ``objective reality'' \ldots might be driven from physics. As we have exhaustively shown here, this anxiety is groundless \ldots'. Heisenberg evidently changes his mind and in Ref.~\cite{HPP} he writes: `Certainly quantum theory does not contain a genuine subjective feature, it does not introduce the mind of the physicist as a part of an atomic event.' But it is the first of these contrasting Heisenberg statements that Wigner takes as justifying the viewpoint underlying the rest of the article:
\begin{quotation}
\noindent
When the province of physical theory was extended to encompass microscopic phenomena, the concept of consciousness came to the fore again: it was not possible to formulate the laws of quantum mechanics in a fully consistent way without reference to consciousness.\end{quotation}

At the heart of Wigner's argument is a rather artificial model (Ref.~\cite{wigner}, p.289) and  here is what he concludes from it:
\begin{quotation}
\noindent
It is at this point that consciousness enters the theory unavoidably and unalterably.\end{quotation} It is soon after this that his friend enters the fray:
\begin{quotation}
\noindent
It is natural to inquire about the situation if one does not make the observation oneself but lets someone else carry it out \ldots One could attribute a wave function to the joint system: friend plus object, and this joint system would have a wave function also after the interaction, that is after my friend has looked. \end{quotation}
It is legitimate to pause here and consider `\ldots the joint system: friend plus object': a wave function is a solution to 
Schr\"odinger's equation but the living, breathing, perspiring friend has an indeterminate number of leptons and baryons and has therefore no determinate hamiltonian, hence no determinate Schr\"odinger equation and hence no wave function. While Schr\"odinger's cat fable serves its original purpose, the usual re-tellings of the story omit to mention what we now know about decoherence and entanglement with the environment. 

When `wave function' enters the discussion, as it does centrally in Wigner's account, one may ask: Into what logical categories might an entity styled `wave function' belong? To the category `marks on a page, blackboard etc.'? Yes, it can. To a category `element of a mathematical formalism'? Yes. Solution to Schr\"odinger's equation? Yes.  According to context, to all of them and more: perhaps `wave function' sometimes refers to the entity of which some of the others are representations? Maybe such an entity could collapse; after all, it is characteristic of what is called a `measurement', that `something' that existed, suddenly exists no longer.  Entities in none of the first three categories could be said to `collapse.' So how would we place in a category something that might logically be said by some people to collapse and that could still be referred to as a wave function? Maybe something represented by an element in a formalism and which could perhaps (logically) be said to collapse? I do not think that this matter is clarified where, at one point Wigner writes `\ldots the wave function is only a suitable language for describing the body of knowledge ...' So do we now have a language collapsing? Shortly after that, the wave function becomes `a convenient summary' \ldots can summaries collapse?
It is easy to be too literal minded, especially when reading an article aimed at a popular audience, but then it is also an article that forms a key part of a recent argument~\cite{FR} challenging the consistency of quantum mechanics. Many discussions of the interpretation of quantum mechanics muddy the water as `wave function' moves from category to category. But one thing seems clear: one cannot attribute `a wave function to the joint system: friend plus object' to any acceptable category.

The key role of consciousness pervades the whole of Wigner's article. For example, it is embedded in the argument on pages 292 - 294 of~\cite{wigner} which ignores the vastly different level of complexity between a `friend'  and an atom (taking `simple physical apparatus such as an atom', p. 293,  to mean `atom', since nothing macroscopic would be appropriate). Wigner also specifically assumes a quasi-Cartesian concept of mind --- mind that does not influence the body. It is surely universal experience that thoughts in the mind can influence the body, so perhaps a completely non-standard meaning applies to `mind' in the article; however it is written as though a standard meaning applies. Incidentally, isn't it a misuse of the term to describe as `cat states' systems that share none of the characteristics of cats, such as continuous complex interaction with the environment?

Wigner's friend is sometimes invoked (not by Wigner~\cite{wigner}) in accounts, e.g. Ref.~\cite{FR},  of the Schr\"odinger's cat situation. Recall, that Schr\"odinger's intention was to show that the \emph{exclusively} linear operation of his equation, and the associated unitary transformation of the state, was not acceptable, since that would lead e.g.\ to ridiculous feline situations. As a result we now have have various alternatives such as deBroglie-Bohm (for material particles), or `many worlds' where unitary transformations continue. But mostly we still have the measurement problem, possibly involving collapse models. But what we do not have is superposed dead and alive cat states. The model that is assumed  by most physicists in their working lives is that there is some sort of expansion into the macroscopic world that occurs very long before the possibility of superposed dead and alive cats arises, any more than we ever see superposed alternative tracks in cloud chambers~\cite{mott}. The conclusion then is that the presence of a friend of Wigner in the Schr\"odinger cat situation could make no difference whatever.

In a recent article~\cite{LEB}, Leslie Ballentine confirms that Wigner publicly disavowed the proposition that consciousness causes collapse, CCC.  His `friend' scenario invokes CCC, and consciousness is invoked by Wigner throughout his article. Arguably, it is time for a serious debate as to whether Wigner's friend should ever be invited to serious discussions relating to the interpretation of quantum mechanics.

A true giant of physics like Wigner, with an unassailably great reputation, has every right to contribute to a collection of partly baked ideas. It is a pity that most reprintings of the article omit to mention its context as given in the book's subtitle.   Most readers are apparently unaware of  this context.

\end{document}